\documentclass[twocolumn]{aastex63}

\hypersetup{linkcolor=red,citecolor=blue,filecolor=cyan,urlcolor=black}

\newcommand{\Msun}{M$_\odot$}

\usepackage{relsize}

\received{August 5$^{th}$, 2022}
\revised{??}
\accepted{??}
\submitjournal{ApJL}

\graphicspath{{./}{figures/}}

\shorttitle{GLASS-JWST -- The faintest red sources}
\shortauthors{Glazebrook et al.}

\begin{document}

\title{Early results from GLASS-JWST XV: properties of the faintest red sources in the NIRCAM deep fields.}


\correspondingauthor{Karl Glazebrook}
\email{kglazebrook@swin.edu.au}



\author[0000-0002-3254-9044]{K. Glazebrook}
\affiliation{Centre for Astrophysics and Supercomputing, Swinburne University of Technology, PO Box 218, Hawthorn, VIC 3122, Australia}

\author[0000-0003-2804-0648 ]{T.~Nanayakkara}
\affiliation{Centre for Astrophysics and Supercomputing, Swinburne University of Technology, PO Box 218, Hawthorn, VIC 3122, Australia}

\author[0000-0003-4239-4055]{C. Jacobs}
\affiliation{Centre for Astrophysics and Supercomputing, Swinburne University of Technology, PO Box 218, Hawthorn, VIC 3122, Australia}
\affiliation{ARC Centre of Excellence for All Sky Astrophysics in 3 Dimensions (ASTRO 3D), Australia)}

\author[0000-0003-4570-3159]{N. Leethochawalit}
\affiliation{School of Physics, University of Melbourne, Parkville 3010, VIC, Australia}
\affiliation{ARC Centre of Excellence for All Sky Astrophysics in 3 Dimensions (ASTRO 3D), Australia}
\affiliation{National Astronomical Research Institute of Thailand (NARIT), Mae Rim, Chiang Mai, 50180, Thailand}

\author[0000-0003-2536-1614]{A. Calabr\`o}
\affiliation{INAF Osservatorio Astronomico di Roma, Via Frascati 33, 00078 Monteporzio Catone, Rome, Italy}


\author{A.~Bonchi}
\affiliation{INAF Osservatorio Astronomico di Roma, Via Frascati 33, 00078 Monteporzio Catone, Rome, Italy}
\affiliation{ASI-Space Science Data Center,  Via del Politecnico, I-00133 Roma, Italy}

\author[0000-0001-9875-8263]{M.~Castellano}
\affiliation{INAF Osservatorio Astronomico di Roma, Via Frascati 33, 00078 Monteporzio Catone, Rome, Italy}

\author[0000-0003-3820-2823]{A. Fontana}
\affiliation{INAF Osservatorio Astronomico di Roma, Via Frascati 33, 00078 Monteporzio Catone, Rome, Italy}

\author[0000-0002-3407-1785]{C. Mason}
\affiliation{Cosmic Dawn Center (DAWN), Denmark}
\affiliation{Niels Bohr Institute, University of Copenhagen, Jagtvej 128, DK-2200 Copenhagen N, Denmark}

\author[0000-0001-6870-8900]{E.~Merlin}
\affiliation{INAF Osservatorio Astronomico di Roma, Via Frascati 33, 00078 Monteporzio Catone, Rome, Italy}

\author[0000-0002-8512-1404]{T. Morishita}
\affiliation{Infrared Processing and Analysis Center, Caltech, 1200 E. California Blvd., Pasadena, CA 91125, USA}

\author[0000-0002-7409-8114]{D.~Paris}
\affiliation{INAF Osservatorio Astronomico di Roma, Via Frascati 33, 00078 Monteporzio Catone, Rome, Italy}

\author[0000-0001-9391-305X]{M. Trenti}
\affiliation{School of Physics, University of Melbourne, Parkville 3010, VIC, Australia}
\affiliation{ARC Centre of Excellence for All Sky Astrophysics in 3 Dimensions (ASTRO 3D), Australia}

\author[0000-0002-8460-0390]{T. Treu}
\affiliation{Department of Physics and Astronomy, University of California, Los Angeles, 430 Portola Plaza, Los Angeles, CA 90095, USA}

\author[0000-0002-9334-8705]{P. Santini}
\affiliation{INAF Osservatorio Astronomico di Roma, Via Frascati 33, 00078 Monteporzio Catone, Rome, Italy}

\author[0000-0002-9373-3865]{X. Wang}
\affil{Infrared Processing and Analysis Center, Caltech, 1200 E. California Blvd., Pasadena, CA 91125, USA}


\author[0000-0003-4109-304X]{K.~Boyett}
\affiliation{School of Physics, University of Melbourne, Parkville 3010, VIC, Australia}
\affiliation{ARC Centre of Excellence for All Sky Astrophysics in 3 Dimensions (ASTRO 3D), Australia}

\author[0000-0001-5984-0395]{Marusa Bradac}
\affiliation{University of Ljubljana, Department of Mathematics and Physics, Jadranska ulica 19, SI-1000 Ljubljana, Slovenia}
\affiliation{Department of Physics and Astronomy, University of California Davis, 1 Shields Avenue, Davis, CA 95616, USA}

\author[0000-0003-2680-005X]{G. Brammer}
\affiliation{Cosmic Dawn Center (DAWN), Denmark}
\affiliation{Niels Bohr Institute, University of Copenhagen, Jagtvej 128, DK-2200 Copenhagen N, Denmark}

\author[0000-0001-5860-3419]{T. Jones}
\affiliation{Department of Physics and Astronomy, University of California Davis, 1 Shields Avenue, Davis, CA 95616, USA}

\author[0000-0001-9002-3502]{D. Marchesini}
\affiliation{Department of Physics and Astronomy, Tufts University, 574 Boston Ave., Medford, MA 02155, USA}

\author[0000-0001-6342-9662]{M. Nonino }
\affiliation{INAF-Trieste Astronomical Observatory, Via Bazzoni 2, 34124, Trieste, Italy}

\author[0000-0003-0980-1499]{B. Vulcani}
\affiliation{INAF- Osservatorio astronomico di Padova, Vicolo Osservatorio 5, I-35122 Padova, Italy} 

\begin{abstract}
We present a first look at the reddest 2--5\micron\  sources found in deep  images from the GLASS Early Release Science program. We undertake a general search, i.e. not looking for any particular spectral signatures, for 
sources detected only in bands redder than reachable with the Hubble Space Telescope, and which would likely not have been identified in pre-JWST surveys.
We search for sources down to AB $\sim 27$ (corresponding to 
$>10\sigma$ detection threshold) {\em in any} of the F200W to F444W filters,with a $>1$ magnitude excess relative to F090W to F150W bands. 
Fainter than F444W$>25$ we find 56 such sources of which 37 have reasonably constrained spectral energy distributions to which we can fit
photometric redshifts. We find the majority of this population  ($\sim$ 65\%) as $2<z<6$ star forming low-attenuation galaxies that are faint at rest-frame ultraviolet-optical wavelengths, have stellar masses $10^{8.5}$--$10^{9.5}
$\Msun, and have observed fluxes at $>$2\micron\  boosted by a
combination of the Balmer break and  emission lines. The typical implied  rest equivalent widths are $\sim$200\AA\ with some extreme objects up to $\sim$1000\AA. 
This is in contrast with brighter magnitudes where the red sources tend to be
$z<3$ quiescent galaxies and dusty star forming objects. 
Our general selection criteria for red sources allow us to independently identify other phenomena as diverse as extremely low mass ($\sim 10^8$ M$_\odot$) quiescent galaxies at $z<1$,  recover recently identified 
$z>11$ galaxies and a very cool brown dwarf.
\end{abstract}


\section{Introduction}


The development of sensitive near-infrared areal detectors for astronomy led to the 
first sky surveys \citep{Gardner1993,gratuitous} and the uncovering of new populations of high-redshift sources. 
The first large area imaging surveys discovered new 
populations of red objects, referred to early on as  
`Extremely Red Objects' or `Distant Red Galaxies'\citep{Pat2004,Franx2003}; contrasting
with the dominant population of `Faint Blue Galaxies' \citep{FBGs}. 
These redder objects
were bright in the near-infrared but dim or undetected in the optical bands. These were later spectroscopically confirmed as mixture of $z\sim 2$ early type massive quiescent galaxies \citep{GDDS,K20,GNIRS}, and massive dusty star-forming galaxies \citep{Wuyts2009}. These populations have now been photometrically and spectroscopically tracked to $z\sim 4$ \citep{Marchesini2010,Spitler2014,S14,Marsan2015,G17,S18,Forrest2020}. The effects of quiescence, dust and redshift all add to make spectral energy distributions (SEDs) progressively redder in the optical to near-infrared bandpasses. In recent years, 
surveys have detected red $H-K$ and $H-3.6\micron$ sources that are likely even higher redshift quiescent and/or dusty sources  \citep{Merlin2019,Fudamoto2021,Marsan2022}.

In the near-infrared the deepest surveys today come from the Hubble Space Telescope, however this is limited to wavelengths $<1.6$\micron.
The state-of-the-art at longer wavelengths has been provided by the 85cm Spitzer Space Telescope which was retired in 2020. Now this is surpassed 
by new data from the James Webb Space Telescope \citep[JWST;][]{Rigby2022} which has unprecedented capability at 2–5\micron\ with the NIRCAM \citep{NIRCAM}
camera and 5--28\micron\ with the MIRI camera \citep{MIRI}. Thus a first look at the sources that emerge in the longer wavelengths of JWST is a compelling prospect. In this paper we do this, utilising data from the GLASS Early Release Science program \citep{Treu2022} where parallel
imaging with NIRCAM provides extremely deep data at $2–5\micron$, and our aim is to characterise the spectral energy distributions and possible nature and redshifts of these sources. In particular we adopt a complementary approach from other
early JWST papers 
(\citealt{Castellano2022}, Paper III; \citealt{Leethochawalit2022}, Paper X;  \citealt{Fink2022,Atek2022,Donnan2022,Naidu2022,Yan2022});
instead of searching for known classes of sources with particular color signatures we use a more general method which is sensitive to a wide variety of sources, and characterise what is revealed by the redder NIRCAM
bands. 

The plan of this paper is as follows: In \ref{data} we describe the data and introduce the general method we use to select red sources. In section \ref{method} we outline our analysis methodology including determination of redshifts, spectral types and stellar masses.
In section \ref{population} we discuss the nature of the population and their spectral energy distributions and likely redshifts. In section \ref{conclusion} we present conclusions. Throughout this paper we adopt AB magnitudes and a standard cosmology with $\Omega_{\rm m}=0.3$ $\Omega_{\Lambda}=0.7$ and H$_0$=70 km s$^{-1}$ Mpc$^{-1}$. 

\section{Data and sample selection}
\label{data}

GLASS-JWST is one of 13 Early Release Science programs. It obtained NIRISS and NIRSpec spectroscopy in the center of the 
massive $z=0.31$ galaxy cluster A2744 on 28--29$^{\rm th}$ June 2022, while obtaining NIRCAM images of two parallel fields 3--8 arcmin away from the 
cluster center. GLASS-JWST consists of the deepest extragalactic data amongst the ERS programs. Details can be found in the survey paper \citep{Treu2022}.
For this paper we consider the NIRCAM parallel fields which are sufficiently distant from the cluster that
only modest lensing magnification is expected \citep{Medezinski2016}. In this paper we neglect the effect, which does not affect colors, and the issue will be revisited after the completion of the campaign. The reduction of the images and construction of photometric catalogs were originally described in  \citet[][Paper II]{Merlin2022}, in this paper we have updated to the Stage 1 data release, with post-flight
calibrations, of \cite{Stage1}. We define the area by the F090W filter coverage, which is only in GLASS, this gives us 
seven JWST filters covering 0.9––4.4 \micron\ over an area of 13.0 arcmin$^2$, with exposures of 1.6--6.5 hours, with the F444W filter being the deepest. The Stage 1 catalogue also contains HST optical photometry, which we use to supplement our SED modelling below. Each object
typically has optical photometry in two HST filters (from coverage in F606W, F775W and F814W Advanced Camera for Surveys filters).

Our catalogue is F444W selected; the F444W image is the detection image and forced photometry is done in the other bands on images PSF-matched to F444W. We correct all bands to total based on the ratio of total to aperture flux in F444W. For this paper's flux 
and color measurements we use an aperture of 0.45 arcsec (this is 3$\times$ the point spread function - PSF - full width half maximum in F444W). The 
5$\sigma$ limiting flux in F444W for this aperture is 28.2, while the other six JWST bands range from 27.9 to 28.4. There are 9525 objects in the catalogue.

We aim to develop a general method to identify sources whose fluxes rise up in the redder bands. First we define the latter:
for `red bands' we utilise the F200W, F277W, F356W and F444W filters. 
Technically F200W is in the NIRCAM `short wavelength' channel but for our purposes we include it in the `red band' category as it  represents a wavelength not accessible to HST and which is limited in depth by considerable thermal emission in ground-based observations. Then the `blue bands' are F090W, F115W and F150W. We require a red selection that picks up a wide variety of
sources and that at the faint end will pick up
objects that are only marginally or not detected in the blue bands, but which at brighter magnitudes can be compared 
with previous HST$+$Spitzer work. After some experimentation we settled on the following:

\begin{enumerate}
    \item We  require that the photometry of a source be good in all 7 NIRCAM bands, i.e. no artefacts or chip boundaries affecting it, which we determined by checking for flagged  pixels near the source center. We also require 2 HST bands, this results in a downselect to 8361 sources.
    \item We define a magnitude we call RED\_BRIGHT, which is the brightest magnitude of a source in {\it any} of the
    red NIRCAM bands.
    \item Next we similarly define BLUE\_BRIGHT for the brightest of the blue NIRCAM bands.
    \item We select sources with BLUE\_BRIGHT $-$ RED\_BRIGHT $>1.0$
    \item We examine the results as a function of the RED\_BRIGHT magnitude limit.
\end{enumerate}

This results in galaxies where at least one of the red bands is one magnitude brighter than {\em all of the blue bands}.
This selection has several advantages: first it can pick up sources that are bright in only one red band (such as might be
due to emission lines contributing at certain wavelengths) as well as continuum sources that are bright in many red bands. The BLUE\_BRIGHT $-$ RED\_BRIGHT $>1.0$ 
selection is defined in AB magnitudes, which is convenient as blue continuum sources such as star-forming galaxies have
$\sim$ constant AB magnitudes with wavelength, and our survey sensitivity is also $\sim$ constant
between bands
(within a factor of two) in Janskies. Secondly 
by utilising a one magnitude break the red color selection is similar to previous methods that have be used to
find high-redshift quiescent galaxies (e.g. \citealt{S14}), dusty galaxies \citep{Marchesini2010,Spitler2014,Franx2003} and Lyman break galaxies \citep{Steidel2003}. Finally at the faint
magnitudes it picks up sources undetected in the blue bands while at bright magnitudes it picks up previously known
red populations.

We consider sources down to RED\_BRIGHT $< 27.0$. At this magnitude limit the peak red fluxes in our aperture are $>10\sigma$, which are robust
sources. Also critically the blue limit for the faintest sources then corresponds to a $>3 \sigma$ detection, so we can be 
confident that the sources are reliably at BLUE\_BRIGHT $-$ RED\_BRIGHT $\gtrsim$ 1 even if
not detected in the blue bands. One caveat to note is
that by construction our catalog is F444W selected, with a point source completeness limit of 29.1 (Paper II). This translates
to $\simeq$ 27.5 for our aperture. Thus although a candidate may be bright in another red band it will always have
some significant F444W flux. An advantage of F444W selection is that it probes
out to $z=7$ the rest frame optical  where
stellar mass-to-light ratios have smaller variation than in the rest frame ultraviolet. We ran a set of simple simulations (following the methodology of \citealp{KG2004} but with $z_{form}=30$) 
using PEGASE.2 models \citep{PEGASE.2} and determined, that for maximally old galaxies, in the absence of significant amounts of dust obscuration, this corresponds to
a strict stellar mass completeness limit of $5\times 10^8$ M$_\odot$
at $z=3$ and $2\times 10^9$ M$_\odot$ at $z=7$. Younger galaxies will be selected below
these mass limits as they have lower mass-to-light ratios.

\section{Methodology}
\label{method}

We use the Stage 1 catalogue and 
applying our BLUE\_BRIGHT$-$RED\_BRIGHT$>1$ selection we obtain 292 
sources with  $RED\_BRIGHT<27$.
Visual inspection of this sample led to the
removal of 76 sources that were associated with image artefacts, blending with
bright neighbours or chip edges. (We note a particularly large cluster of these around a 17$^{\rm th}$ magnitude star.)
This gives a sample for analysis of 216 sources.

For these we fit the photometric redshifts and SEDs using the EAZY software \citep{EAZY},
specifically {\tt eazy-py} version 0.5.2.  
Our EAZY fits use the new template set of \cite{Larson22} which include high
equivalent width emission line components which have proved important for fitting high-redshift JWST sources.
EAZY is a robust and accurate photometric redshift
and multi-component SED fitting tool
that has been utilised and validated in many deep surveys (e.g. \citealt{S16,Whit11,3DHST}).
A comparison of EAZY redshifts with spectroscopic redshifts in the GLASS fields \citep[PaperXVI]{PaperXVI} shows good performance
with 2\% redshift accuracy,
however in our new regime of the faintest JWST objects we approach this with caution.

To assess photometric redshift performance we consider the probability distributions $p(z)$ returned by
the EAZY fits as internal error estimates. We derive lower and upper redshift bounds, and a redshift error $\Delta z$,
 by calculating a 68 percentile interval around the
best fit redshift. We then define  `good' photometric redshifts as those where the $p(z)$ gives $\Delta z/(1+z)<0.5$.
For bright objects ($RED\_BRIGHT\le 25$) this internal redshift accuracy is good, we find that the median
error on $\Delta z/(1+z)$ is 0.05 with 98\% good fits. For fainter objects ($25< RED\_BRIGHT\le 27$, see below for this choice) we see a mix of compact and broad $p(z)$ curves. We find good fits have a median
$\Delta z/(1+z)=0.12$ and constitute 67\% of the faint sample.  Visually the $p(z)$ plots are compact with single peaks and weak or absent secondary peaks. The `bad' photometric redshifts have broad $p(z)$ curves often without prominent peaks. Selecting 
on good photometric redshifts results in 155/158 objects in the bright sample and 39/58 in the faint sample.
We exclude the bad photometric redshift sources from further analysis.
We note the photometric redshift performance
of our selected sample may be different than that of the general population as we have selected
objects with strong color signatures.

\begin{figure*}
\setcounter{figure}{0}
\plotone{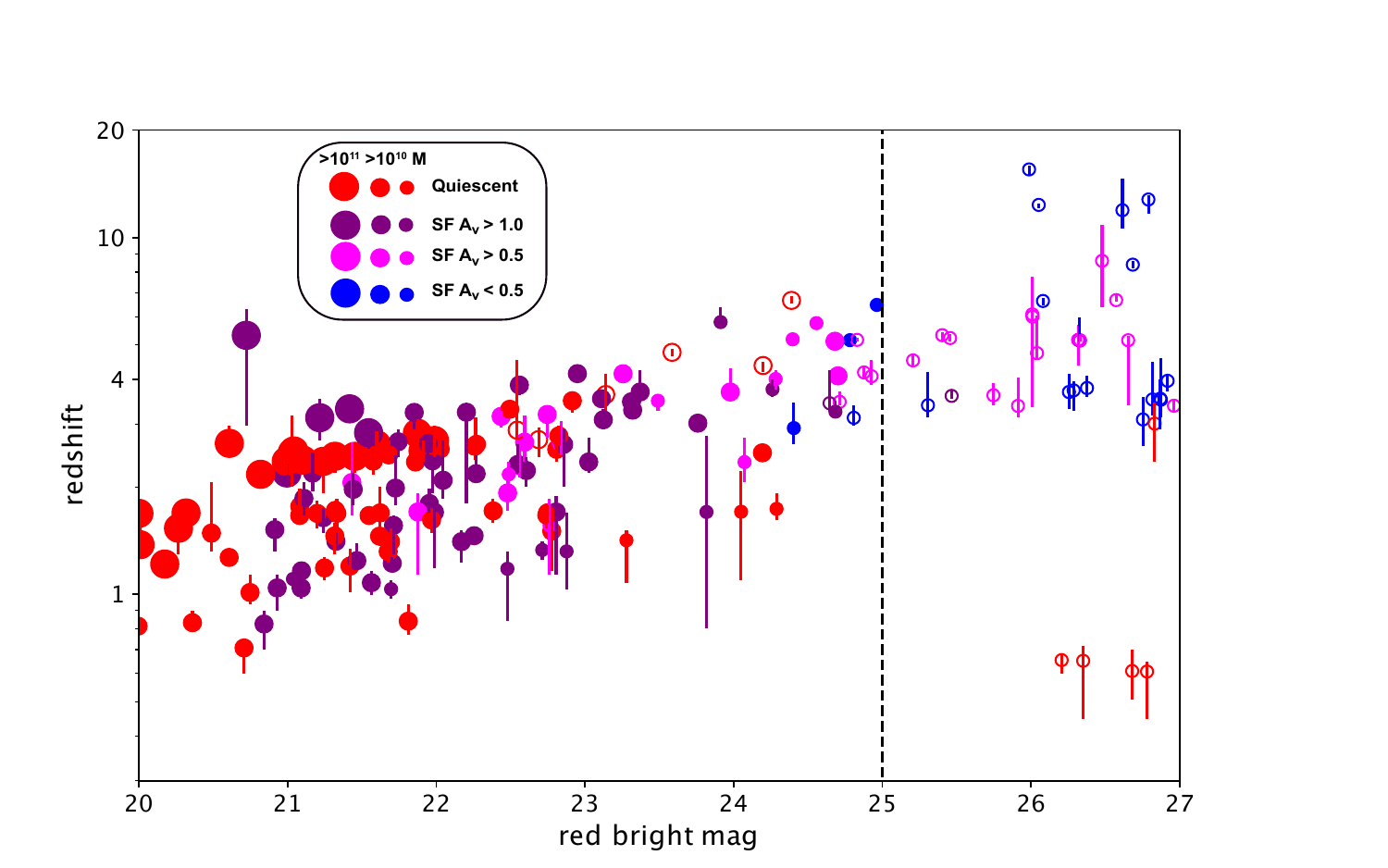}
\caption{Redshift--magnitude distribution of the 
BLUE\_BRIGHT$-$RED\_BRIGHT $>1$ selected  sources. Vertical bars denote the redshift uncertainty. Symbol size and
color are keyed to stellar mass, spectral type (quiescent/star forming) and dust attenuation 
as given in the legend. Open circles are the same coding, but denote 
objects that would be undetected in F160W in a deep HST survey, while the dashed vertical line shows the approximate Spitzer confusion limit. 
}
\end{figure*}

To derive indicative physical properties we then do  further SED fits, assuming the photometric redshifts, using 
Prospector \citep{Johnson2021a}. This allows us to obtain stellar masses, 
star formation histories
and dust attenuation values. Prospector includes a greatly improved physical treatment of complex
star-formation histories and the effect of emission lines on the
photometry. 
We use a non-parametric  {\tt continuity\_flex\_sfh}  with 4 SFH bins. 
We use a \citet{Kroupa2001} Initial Mass Function and fix the redshift of the galaxies at the best fit EAZY values. 
We use a \citet{Calzetti2000} dust law and let the dust optical depth vary between  0--2.0. 
We vary the stellar metallicity between $\log_{10}(\mathrm{Z}/\mathrm{Z}\odot)=-2$ to 0.19.
We further fix gas phase metallicity to be same as stellar metallicity and allow the ionisation parameter of the galaxies to vary between U=$-1$ to $-4$. We have inspected the Prospector SED 
fits and find them to agree well with the EAZY SED fits. Three of the sources (one with $RED\_BRIGHT<25$)  had Prospector fits that failed to converge, we removed these from the
sample for further analysis.
We included redshift errors in the Prospector analysis by re-fitting the SEDs at the 68 percentile upper and lower redshift bounds derived above. We then merge the upper and lower limits of the physical parameters across this redshift range.
These derived quantities for the sample of 216 sources is given in Table~1.

In Figure~1 we plot RED\_BRIGHT vs photometric redshift for our sources and 
mark the typical limits of HST and Spitzer surveys. As a reference for this we take the Hubble Frontier Fields (HFF) depth from \cite{Shipley2018} which is 
the deepest near-infrared survey with HST. Their HST F160W point source completeness 
limit when corrected for our aperture corresponds to BLUE\_BRIGHT$=26.0$, we mark objects fainter than this in the blue channels with open circles. 
The Spitzer 3.6$+$4.5\micron\ bands are similar to our F356W and F444W bands. In the HFF their
depth was AB$=$25 (an
aperture correction is inapplicable as Spitzer's broad PSF makes faint objects 
effectively point sources). While there are significantly deeper Spitzer surveys they become
seriously confusion limited and incomplete for AB$>25$ (see Figure~14 of \citealt{S-CANDELS}). This issue is normally addressed by modelling Spitzer fluxes using
HST images as priors on source location, this introduces a dependence on detection in the bluer bands.
Therefore we mark RED\_BRIGHT$=$25 as the approximate limit for sources found with Spitzer, 
noting that forced Spitzer photometry of HST detected sources can go considerably deeper.

We define `quiescent galaxies' as those with  $\log_{10}$ of the
specific star-formation rate per year (hereafter {\tt logssfr}) as $<-9.4$. This is a factor of 4 
below the main sequence at $3<z<4$ from \cite{S18}. 
 We estimate dust attenuation $A_V$ from the {\tt dust2} parameter of the Prospector SED fits and code this in 3 bins on Figure~1. By inspecting the SEDs by eye we have verified that these attenuation classifications
accord well with the shape and steepness of the best fit SEDs.

\begin{figure*}
\setcounter{figure}{1}
\centering
\includegraphics[width=18cm]{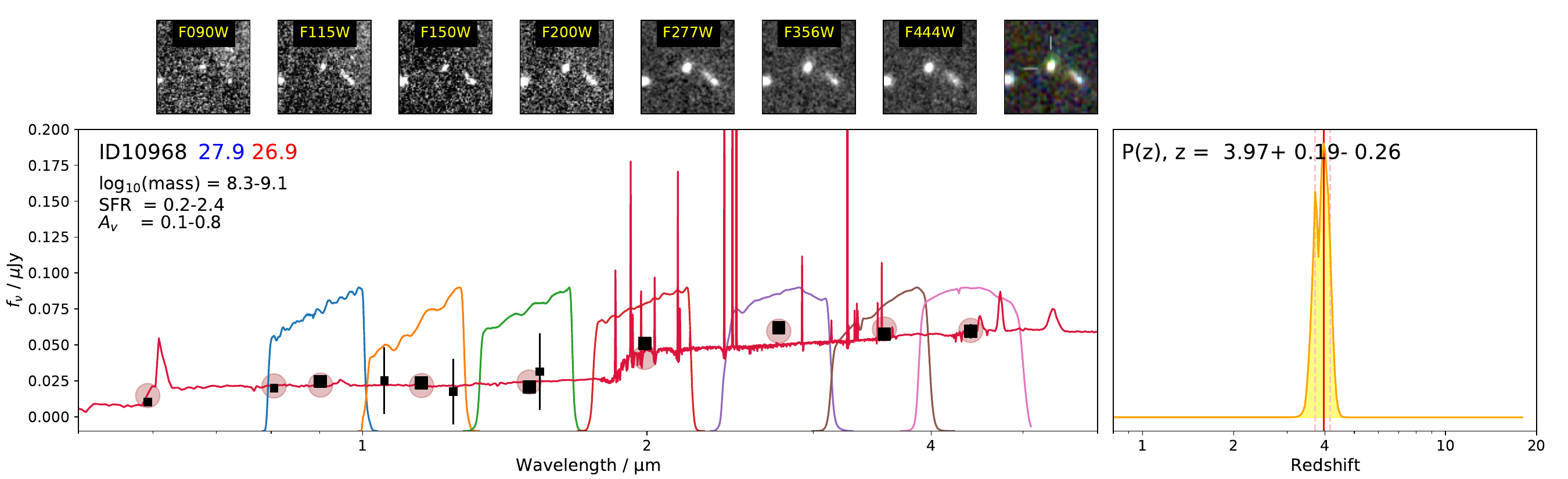}
\includegraphics[width=18cm]{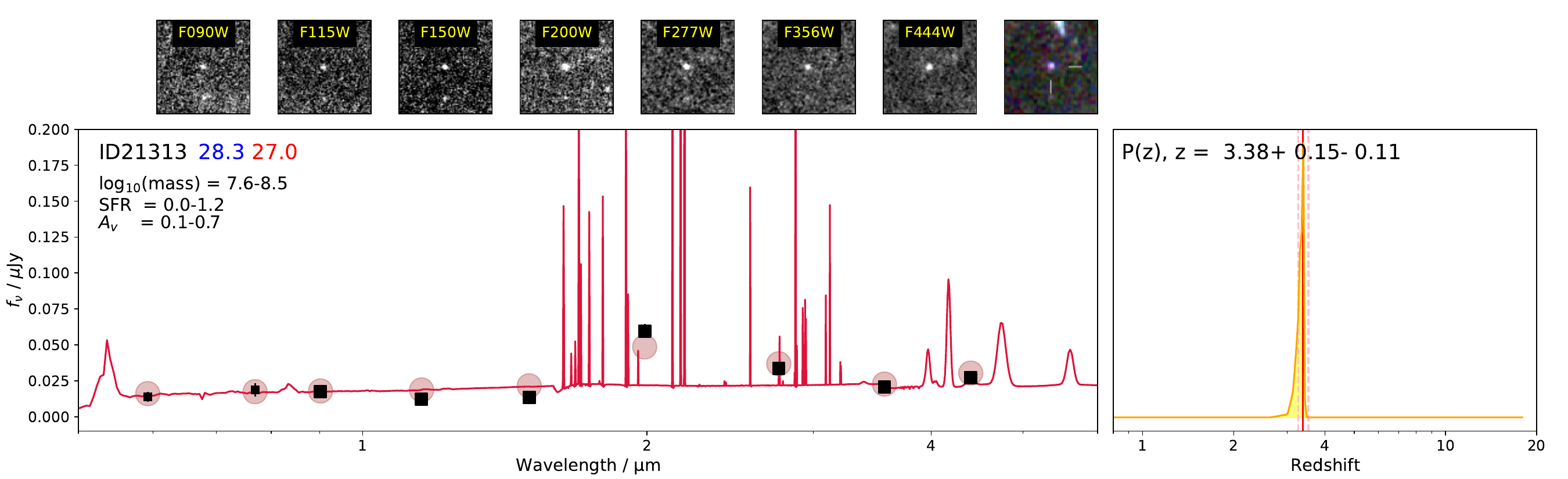}
\includegraphics[width=18cm]{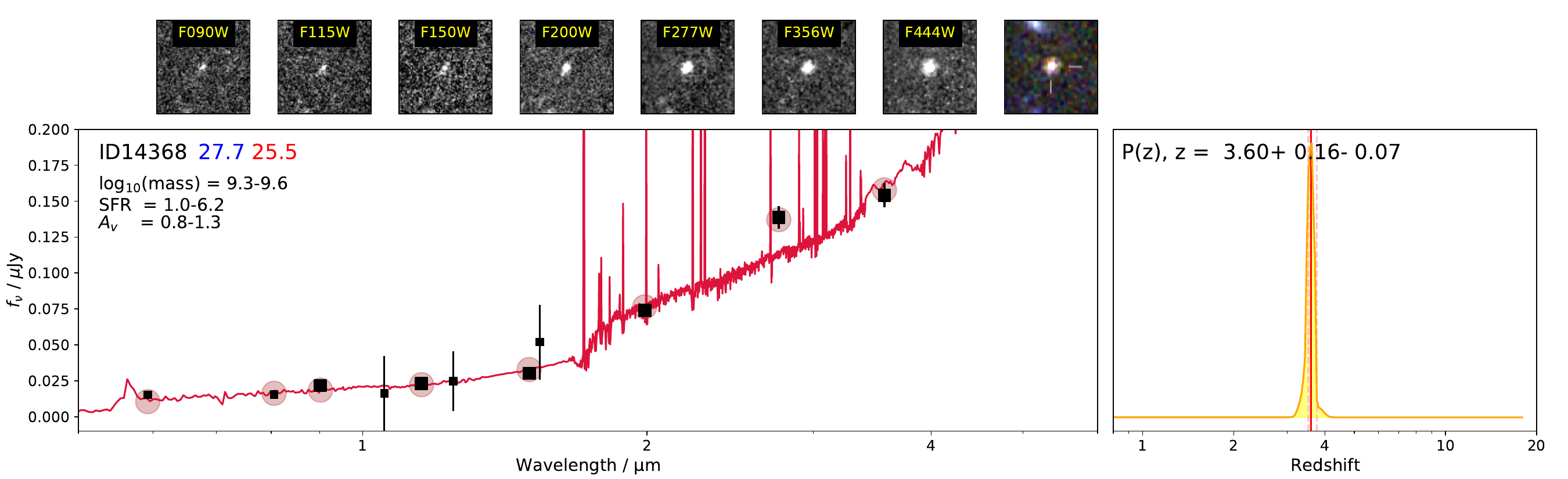}
\includegraphics[width=18cm]{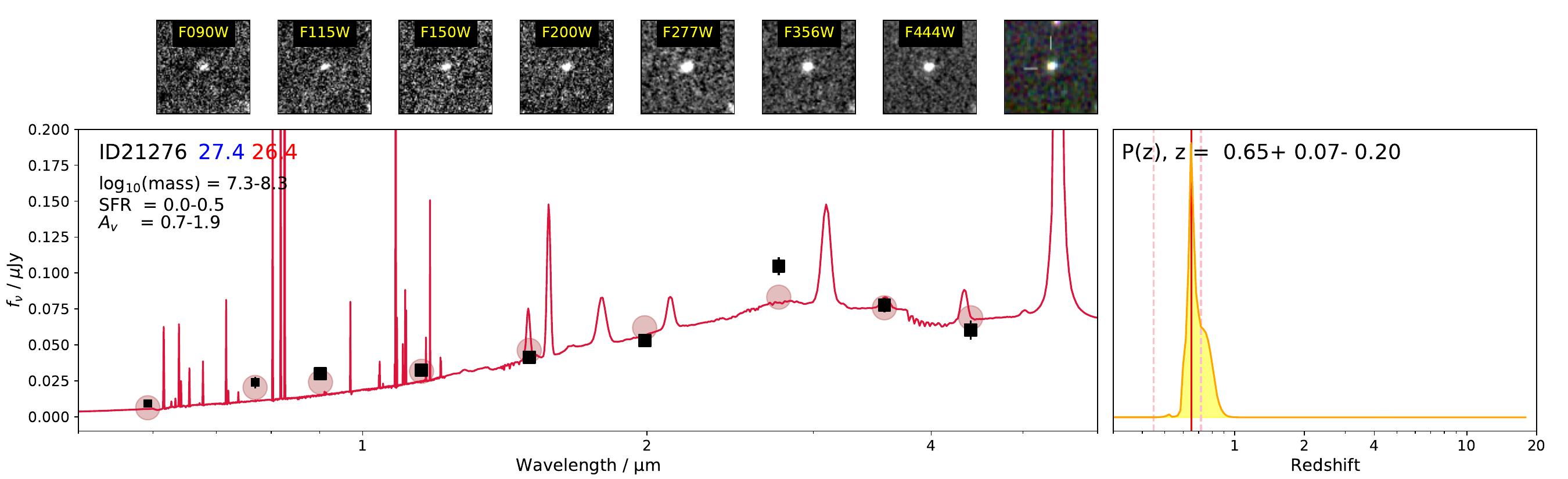}
\caption{Image montages and Prospector SED fits of example objects. As well as the
individual bands we show a `wide' RGB color image constructed from F444W, F356W and
F200W. The black points in the SED plots show the observed photometry with the larger symbols being
the GLASS NIRCAM bands,
the red line shows the best SED from Prospector. The colored text shows the corresponding
BLUE\_BRIGHT and RED\_BRIGHT magnitudes. The large red points
are photometry of the best fit SED  and the legend shows physical parameter's 68\% range. 
$p(z)$ probability distributions are shown for the SED fits in the right panels with vertical lines
denoting best fit redshifts and uncertainties.
In the top plot the NIRCAM filter transmission profiles are plotted to show their extent.
}
\end{figure*}

\section{Discussion of Sources}
\label{population}
Several trends are apparent in the source population. First it can be seen 
in Figure~1 that at bright magnitudes the sample is dominated  by
quiescent galaxies and dusty star forming galaxies at $z\sim 2$. This is a well known result as discussed
in the introduction; and one might expect to see more such things at fainter magnitudes. However the nature of the population shifts and we see that at RED\_BRIGHT$>$25 the population is dominated ($\sim$ 65\%) by low attenuation ($A_V<1$)
star forming galaxies at $2<z<6$. From the Prospector fits we find the typical stellar masses are $10^{8.5}$--$10^{9.5}
$\Msun, with a typical error of 0.2 dex.
We also see candidate star-forming galaxies at $z>11$ appearing, which we will discuss in detail below.


We present examples showing the ranges of sources at the faint end in Figure 2. ID numbers refer
to the Stage I catalog. 
To start with ID10968 and ID21313 show examples of blue $z\sim 4$ star forming galaxies that are the dominant population of galaxies selected by our criteria. It can be seen that the increased flux $>2$\micron\ comes from  the Balmer break together with a  contribution from H$\beta$ and [OIII] emission lines.  ID10968 has a pronounced Balmer break. In contrast ID21313 has a very large contribution from emission lines which is
pronounced in the F200W filter, this is evident by eye in the image. In general we find that many SEDs can not be fit without
a strong
line flux contributions, if one removes  high equivalent
width templates from EAZY then the median $\chi^2$ SED residual of the faint sample increases significantly from 7 to 13. 

\begin{figure*}
\setcounter{figure}{1} 
\centering
\includegraphics[width=18cm]{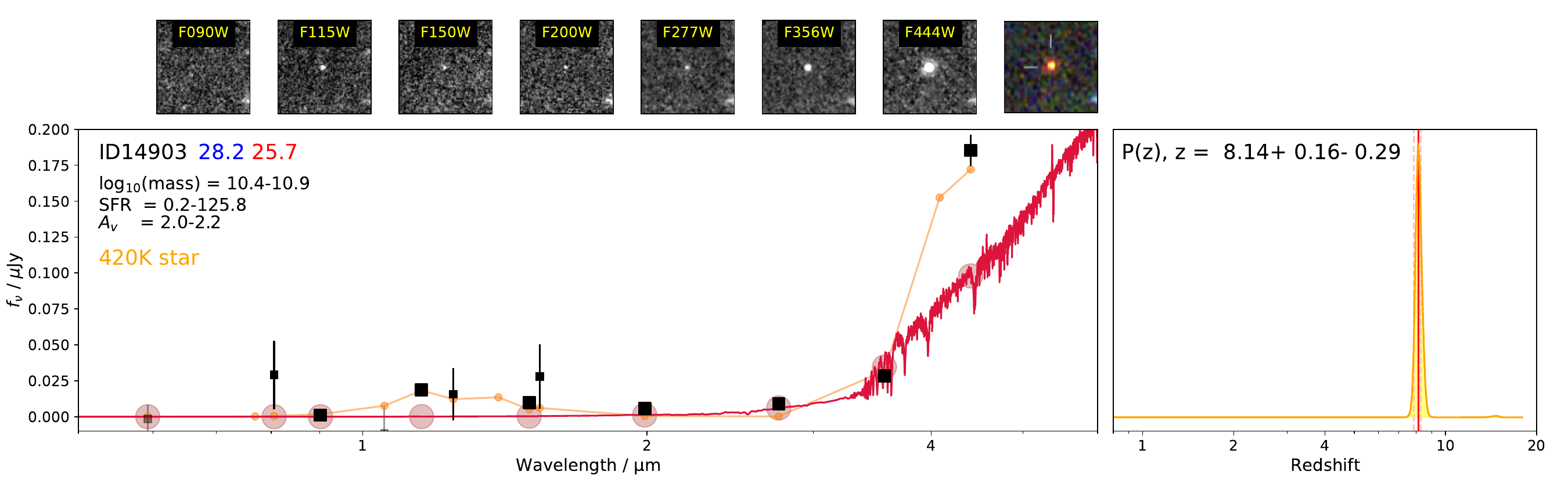}
\includegraphics[width=18cm]{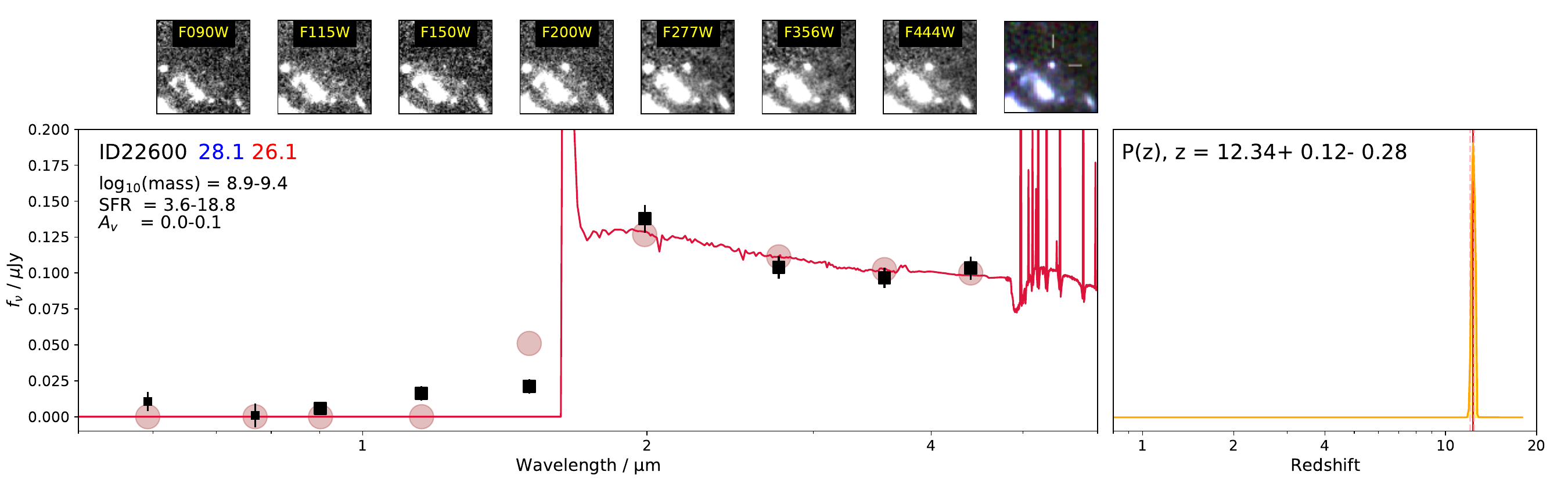}
\includegraphics[width=18cm]{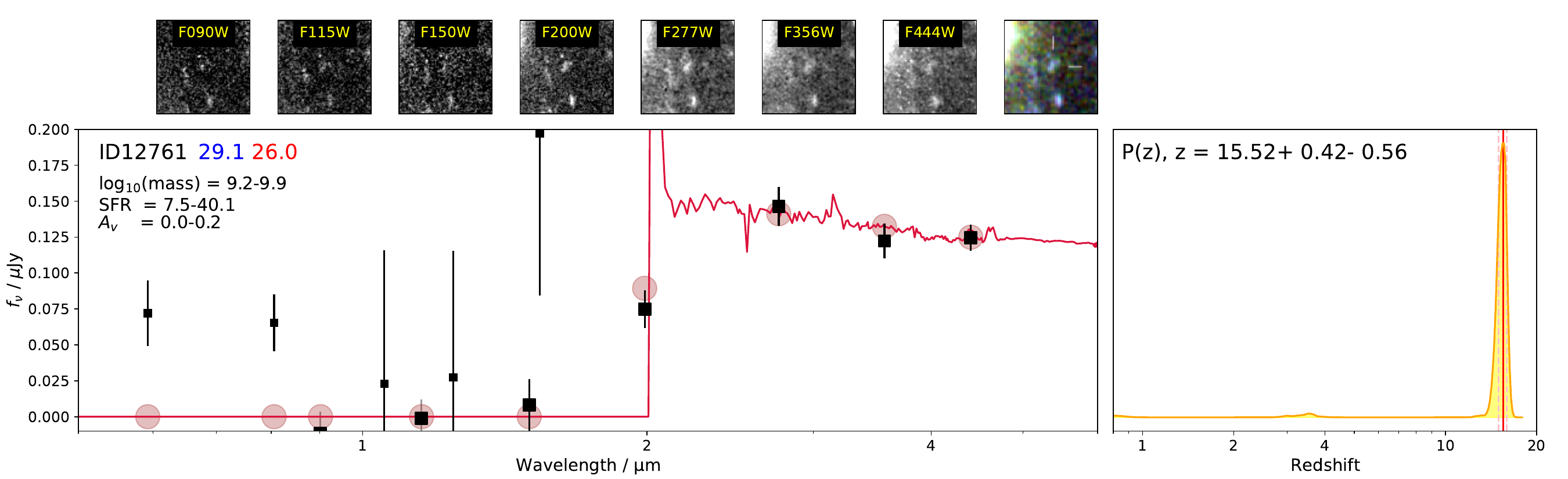}
\caption{{\bf (contd.)} Image montages and SED fits of further sources.
ID14903 agrees much better with an ultra cool stellar template 
(shown as the orange curve) rather than a high-redshift galaxy.
We show two  $z>11$ Lyman break
galaxy candidates. ID22600 is a strong candidate \citep{Castellano2022},
ID12761 is a poor candidate and likely contaminated despite the strongly peaked $p(z)$.
}
\end{figure*}

The need for emission line contributions  was notable, so we investigated what level of emission line equivalent widths
were needed to give such boosts to the photometry. To do this we measured the summed
H$\beta$ + [OIII] 4959,5007\AA\ equivalent widths of the best fit Prospector models. We estimate the errors
on this equivalent width as:
$$ \Delta  \hbox{EW}  = \frac{\Delta f_\nu}{df_\nu / d\hbox{EW}} $$
where the derivative  $d f_\nu / d \hbox{EW}$ is estimated from the best fit model using the fractional contribution of the
emission lines to the flux $f_\nu$ in the nearest NIRCAM band, with $\Delta f_\nu$ being the photometric error.

As an indication of how
`blue' the galaxies might be we also calculate the rest frame ultraviolet continuum slope $\beta$ following 
the method of PaperXVI.

We plot equivalent width against $\beta$  Figure~3. The median equivalent width is 180\AA, this is high compared to $z=0$ but is typical
for galaxies of these masses at $z\sim 4$ (compare for example Figure~3 of \cite{Reddy2018}. Similarly the
$\beta$ values are also consistent with previous measurements of normal star-forming galaxies at this redshift \citep{Bouwens2014,Reddy2018}.
In particular we see no extremely blue values ($\beta<-2.5$). 

In our sample there are a handful of galaxies with equivalent width $>400$\AA\ (for example ID21313 in Figure~2 is $1370 \pm 150$\AA). These objects not surprisingly
have the highest specific star formation rates, with {\tt logssfr}$\sim 8.2$ or $6\times$ the median value.
Their space density is $\sim 3\times 10^{-5}$ Mpc$^{-3}$ which comparable to that found using medium band 
filters by \citet{Forrest2017}.

ID14368 shows an example of a dustier star forming galaxy at $z\sim 3.6$ with $A_V=1.1$, these are less common in the faint sample. 
Examples of even rarer selected sources are shown on the lower panels. ID21276
shows a quiescent galaxy candidate (star formation rate $< 0.03$ M$_\odot$ yr$^{-1}$  at $z=0.7$
 with an extremely low stellar mass of $\sim 10^8$ M$_\odot$.
 and with moderate dust attenuation ($A_V=1.7$). This is below the completeness
 limit of stellar mass functions determined from deep ground near-infrared surveys
 \cite{Tomczak2014}. We note it has F200W $=$ 27.1, considerably below the limit of ground based $K$-band surveys \citep{S16}.
In a companion paper (Paper IX; \citealt{QGs}) we present
the first spectra from JWST of two low mass ($\sim 10^{10}$ M$_\odot$) $z\sim 2$ quiescent galaxies. These results
augur well for the future prospects of JWST
to measure the properties of quiescent galaxies at low masses.

ID14903 is a point source and has an unusual SED with a strong rise between F3456W and F444W; the residual flux in F115W strongly rules
out a $z\gtrsim 8$ solution. The galaxy fit is poor. It is much better matched by a cool star SED, using the Phoenix stellar templates
built in to EAZY we find a 400K Y dwarf is an excellent fit. This 
demonstrates how important it is to consider cool star templates
when evaluating very high redshift solutions.
We explore this object in more
detail in our companion Paper XIII \citep{T-dwarf} -- which describes the independent discovery -- with a more sophisticated set of
stellar templates and conclude it is a star on the T/Y boundary. It is the first ultra cool
dwarf to be discovered by JWST, its faint magnitude places it well outside the Milky Way thin disk.

ID22600 is a high confidence $z=12.3$ Lyman break galaxy candidate with a pronounced Lyman dropout between F150W and F200W, this was presented in detail in our
companion Paper III \citep{Castellano2022} where it was discovered by classical Lyman break color
selection. We note the other bright galaxy in that paper at $z=10.6$ is too low redshift to be selected by our method here; it has too much flux in F150W. Our method is not sensitive 
to Lyman break galaxies with redshifts $7<z<11$ as they have strong rest-ultraviolet
continuum in the blue bands.


\begin{figure}
\setcounter{figure}{2} 
\centering
\includegraphics[width=8cm]{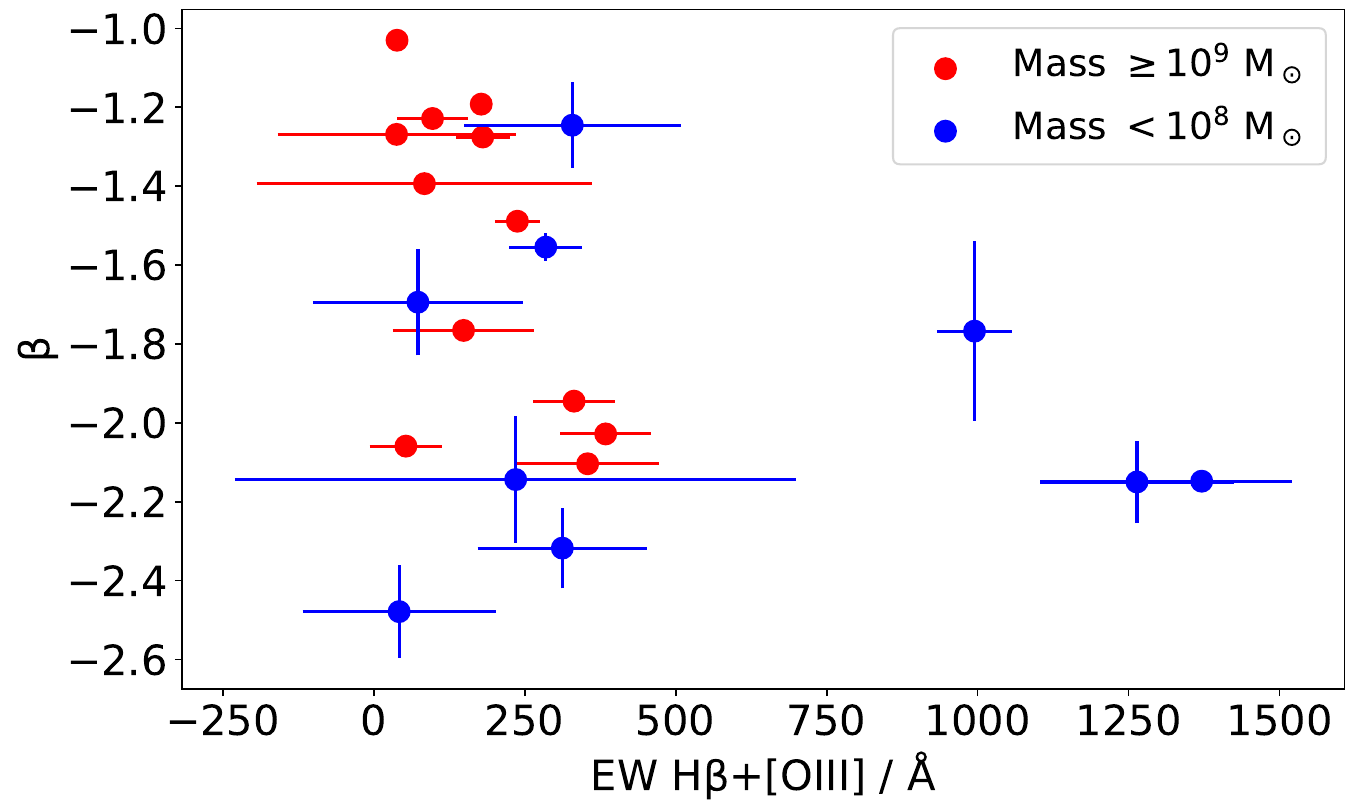}
\caption{ Photometrically inferred emission line (H$\beta$ + [OIII] 4961,5007\AA) equivalent widths 
and rest-frame ultraviolet $\beta$ values for 
objects in our selected sample with RED\_BRIGHT$>$25 and $2<z<6$. }
\end{figure}

There are three more $z>11$ candidates in our selection; however none are compelling. ID12761 shows
the $z=15.5$ candidate which is the highest redshift one. However there is weak residual flux shortwards
of $1\micron$ which rules this out. The photometry at long wavelengths may be contaminated
by a nearby bright galaxy.   The other two objects have similar photometric issues. 
The discovery of $z\sim 16$ F150W dropouts has attracted a lot of recent attention \citep{Fink2022,Donnan2022,Atek2022} and is scientifically
important for our understanding of early galaxy formation, however as see here
SEDs at these redshifts may be ambiguous unless they have very high signal:noise
\citep[e.g.,][]{Zavala2022}. Future improved GLASS reductions will allow these candidates
to be reanalysed, and they could be targeted for future JWST spectroscopy along with more
firm candidates.
These results do however indicate that our technique is a promising 
alternative to traditional methods to discover more of the very high-redshift objects.

\section{Conclusions}
\label{conclusion}

We make a first exploration of the deep sky considering the faintest
very red sources that emerge at wavelengths $>2$\micron\ in JWST NIRCAM bands. Such sources would not have been seen by previous surveys. We utilise a novel general search method
that does not depend on any particular choice of SED class to search for.
We find 56 such faint sources ($\sim 4$ arcmin$^{-2}$) that are detected in one or more bands beyond $2\micron$ but are absent or only marginally detected in bluer bands. We are able to make assessments of the nature of 37 of them. Our primary conclusions are:

\begin{enumerate}
\item Our novel selection method picks out a diversity of different classes of interesting sources.

\item Contrary, perhaps, to a naive intuition, the population is dominated by low mass faint blue galaxies
at $z\sim 4$, where the Balmer break and strong H$\beta+$[OIII] emission lines are redshifted into
the red bands.

\item We find a few exotica such as a cool and distant T dwarf star and very low mass quiescent galaxies at $z<1$.  

\item We recover a robust $z=12.3$ Lyman break galaxy found by earlier color selection and
identify  additional, weaker, candidates at $z>11$. However, these two are not robust, with evidence
of contaminating flux. Nevertheless, this shows that our method 
has the potential to be a useful alternative to classical techniques in such searches.

\end{enumerate}

This initial study -- selecting objects ranging from extreme line emitters, passive galaxies, galaxies at $z>11$ to brown dwarfs, demonstrates the power of JWST, in particular the red channels beyond HST limits,  to discover and characterize new astronomical phenomena. This
analysis is only a preliminary first look to see what is revealed by red NIRCAM channels.  Future work can greatly improve the statistics utilising future improved NIRCAM calibrations and deeper and wider JWST surveys. NIRCAM slitless spectroscopy will be able to quickly make a detailed
census of $z\gtrsim 3$ line emitters and NIRSPEC will be powerful for confirming sources at all redshifts considered.
Finally, it would be valuable to add mid-infrared data from MIRI to better characterise the full SED shapes of the reddest objects that JWST/NIRCAM will find.

\acknowledgments

This work is based on observations made with the NASA/ESA/CSA James Webb Space Telescope. The data were obtained from the Mikulski Archive for Space Telescopes at the Space Telescope Science Institute, which is operated by the Association of Universities for Research in Astronomy, Inc., under NASA contract NAS 5-03127 for JWST. These observations are associated with program JWST-ERS-1324. We acknowledge financial support from NASA through grants JWST-ERS-1342. 
KG, TN and CJ acknowledge support from Australian Research Council Laureate Fellowship FL180100060. NL and MT acknowledge support by the Australian Research Council Centre of Excellence for All Sky Astrophysics in 3 Dimensions (ASTRO 3D), through project number CE170100013. CM acknowledges support by the VILLUM FONDEN under grant 37459. The Cosmic Dawn Center (DAWN) is funded by the Danish National Research Foundation under grant DNRF140. MB acknowledges support from the Slovenian national research agency ARRS through grant N1-0238.
\bibliographystyle{aasjournal}

\bibliography{myreferences}{}

\begin{deluxetable}{lll}
\tablecaption{Contents of Table 1}
\tablehead{
\colhead{Column} & \colhead{Units} & \colhead{Explanation}
}
\startdata
1 & --- & Identifier  from Paris et al. (2023) photometric catalogue \\
2,3& nJy& The BLUE\_BRIGHT flux value and flux uncertainty  \\
4 & --- & Filter selected for BLUE\_BRIGHT flux \\
5,6& nJy & The RED\_BRIGHT flux value and flux uncertainty   \\
7 & --- & Filter selected for RED\_BRIGHT flux \\
8,9,10 &  --- & photometric redshift  and upper and lower uncertainty\\
11,12,13 & M$_\odot$  & $\log_{10}$ of the stellar mass and upper and lower uncertainty$^\dag$\\
14,15,16 &M$_\odot$ yr$^{-1}$  & Star formation rate and upper and lower uncertainty$^\dag$\ \\
17,18,19 & mag & Dust  attenuation $A_V$ and upper and lower uncertainty$^\dag$\\\
20,21 & \AA & Rest frame equivalent width of H$\beta+$[OIII]4959,5007 and uncertainty$^\dag$\\\
22,23 & --- & Rest frame ultraviolet slope $\beta$ and uncertainty$^\dag$\\\
\enddata
\tablecomments{Table 1 is published in its entirety in the electronic 
edition of the {\it Astrophysical Journal}.  The description of the columns
is given here.\\
\\
$\dag$ Missing values have $-99$ in the uncertainty }
\end{deluxetable}

\end{document}